\documentclass[oribibl]{llncs}

\usepackage{amssymb}
\usepackage{amsmath}

\usepackage[colorlinks = true,
  linkcolor = blue,
  citecolor = blue,
  urlcolor = blue]{hyperref}

\usepackage{url}

\usepackage{fancyhdr} % for arxiv
\fancypagestyle{firststyle}
{
\fancyhf{}
\fancyfoot[C]{\scriptsize{Proceedings of the 31st International Working Conference on Requirements Engineering: Foundation for Software Quality (REFSQ 2025), Barcelona, Springer, pp. 295--303. This version is the authors' copy. The publisher's definite version is available online via \url{https://doi.org/10.1007/978-3-031-88531-0_21}}.}
}

\begin{document}

\title{The Potential of Citizen Platforms for Requirements Engineering of Large
  Socio-Technical Software Systems}

\author{Jukka Ruohonen\inst{1}\orcidID{0000-0001-5147-3084} \and Kalle Hjerppe\inst{2}}
\institute{University of Southern Denmark, S\o{}nderborg, Denmark, \email{juk@mmmi.sdu.dk} \and University of Turku, Turku, Finland, \email{kphjer@utu.fi}}

\maketitle

\begin{abstract}
Participatory citizen platforms are innovative solutions to digitally better
engage citizens in policy-making and deliberative democracy in general. Although
these platforms have been used also in an engineering context, thus far, there
is no existing work for connecting the platforms to requirements
engineering. The present paper fills this notable gap. In addition to discussing
the platforms in conjunction with requirements engineering, the paper elaborates
potential advantages and disadvantages, thus paving the way for a future pilot
study in a software engineering context. With these engineering tenets, the
paper also contributes to the research of large socio-technical software systems
in a public sector context, including their implementation and governance.
\end{abstract}


\section{Introduction}

\thispagestyle{firststyle} % for arxiv

In recent years there has been an increasing interest to better engage citizens
in public policy-making due to real or perceived problems in representative
democracy~\cite{Bachtinger18, Barandiaran24, Macq23}. The argument is that
citizens want to engage in politics and democracy, but they do not have
knowledge or means to do so. The increasing digitalization of societies has
allegedly also rendered the traditional participatory infrastructures outmoded;
the old democratic \textit{agoras} of a polity have disappeared or became
dysfunctional. At the same time, large amounts of citizens have moved to social
media platforms within which the many real or perceived problems are
well-recognized; the list includes, but is not limited to, algorithmic
polarization, engagement and amplification, so-called echo chambers and filter
bubbles, disinformation, misinformation, and propaganda, hate speech and
harassment, and many other issues~\cite{Ruohonen24FM}. According to critical
viewpoints, deliberative democracy is in trouble due to these and other related
trends~\cite{McKay21}.\footnote{~In addition to direct democracy, deliberative
  democracy is one of the leading concepts to patch the current problems in
  representative democracy. Drawing on ideas that can be traced all the way back
  to the ancient times, the essence is that ``people come together, on the basis
  of equal status and mutual respect, to discuss the political issues they face
  and, on the basis of those discussions, decide on the policies that will then
  affect their lives'' \cite[p.~2]{Bachtinger18}.}  To this end, these trends
have prompted new arguments and ideas about the difficult relationship between
democracy and its institutions, including knowledge, markets, and the so-called
``marketplace of ideas''. For instance, it has been recently argued that
expertise and epistemology are at the heart of the problems; citizens must have
means to receive knowledge that matters for policy-making, and they must have
functioning democratic channels to subsequently voice their knowledge-based
opinions~\cite{Herzog24}. This argument suits well for framing also the paper's
scope.

Thus, on one hand, expertise and knowledge are essential also in requirements
engineering; here, software or other engineers are the experts who solicit
requirements from non-experts, including customers, clients, and
stakeholders. To further frame the scope, the paper considers requirements
engineering of large socio-technical systems in a public sector context; hence,
the non-experts include also citizens for whom the systems are foremost
implemented. Then, on the other hand, the previous points about platforms frame
the paper's central idea; the potential use of so-called citizen platforms for
requirements engineering in the context described. These platforms have been
successfully used in some countries for participatory policy-making and
deliberative democracy in general.

While there is existing work also on the use of citizen platforms in large-scale
engineering projects, such as so-called smart cities~\cite{Leclercq22,
  Simonofski21}, the paper is the first to contemplate these explicitly in
relation to requirements engineering. The note about smart cities exemplifies
what is meant by large socio-technical systems. Fundamentally, cities belong to
the public sphere, and adding smart technologies to them thus necessitates the
involvement of public sector stakeholders, including citizens, even in case the
technologies themselves are designed, engineered, and operated by private sector
companies. On these notes and clarifying framings, the opening Section~\ref{sec:
  citizen platforms} briefly elaborates what citizen platforms are about. The
subsequent Section~\ref{sec: citizen platforms re} continues by discussing their
potential use in requirements engineering. A conclusion and a discussion are
presented in the final Section~\ref{sec: conclusion}.

\section{Citizen Platforms}\label{sec: citizen platforms}

Citizen platforms are innovative means for better civic engagement in the
current, increasingly digitalized societies. In essence: citizens gather on a
platform for discussing, debating, and deliberating about a given public policy
issue, eventually reaching a consensus everyone can agree upon. Taiwan has been
the leading country in the adoption of citizen platforms; over the years,
numerous platforms have been deployed there and successfully used to
democratically solve policy issues~\cite{Banerjee23}. Recently, citizen
platforms have been developed or deployed also elsewhere, including in
Finland~\cite{Sitra24} and Spain~\cite{Barandiaran24}. As a testimony of the
potential offered, similar ideas and implementations have been adopted for even
more challenging scenarios, such as peace-building and conflict
resolution~\cite{Bell24}. Against this brief background, it seems sensible that
citizen platforms could be used also for requirements engineering in the public
sector context in particular.

These platforms may overcome a long-standing sin in requirements engineering;
the involvement of only a few stakeholders at the expense of users, citizens,
employees, consumers, and other people for whom many systems are
designed~\cite{Dalpiaz16}. Given the public sector context, citizen platforms
may also help at gathering voices from underrepresented or even marginalized
citizen groups. The obvious benefit would be supposedly increased citizen
satisfaction with a given socio-technical system implemented. In addition,
deliberative participation may increase trust placed upon a system, among other
things. Even more importantly, the platforms may catch two important birds with
a single stone; not only may these increase satisfaction with and trust toward a
socio-technical system, but these may also do the same with respect to democracy
itself.

Three clarifications are needed before continuing. First: although the basic
idea is rather similar, citizen platforms should not be confused with so-called
citizen assemblies, which have been increasingly enacted in Europe in recent
years~\cite{Macq23}. Second: although somewhat similar in terms of technical
implementations, citizen platforms are different from requirements engineering
platforms designed for crowd-sourcing contexts~\cite{Dalpiaz16,
  Groen17}. Importantly, citizens in a society are not a some anonymous crowd in
the Internet; they have a right to participate and have their voices heard on
public policy matters affecting them. The existing citizen platforms have also
been strongly influenced by deliberation; participants on a platform seek a
consensus through discussion. They are thus not merely a crowd from whom
feedback is solicited. Third: from a broader perspective, citizen platforms also
interlace with the discussion on public alternatives to the large multinational
social media platforms and commercial platforms in
general~\cite{Winseck22}. This point underlines the public sector context. It
serves also as an argument in favor of citizen platforms; citizens and
stakeholders might be more eager to participate on an open source citizen platform
orchestrated by a public sector body due to privacy, data protection, integrity,
trust, and ethical reasons.

In terms of existing implementations,
Polis\footnote{~\url{https://compdemocracy.org/Polis/}} is the pioneering
citizen platform. The discussions on the platform do not mimic those on social
media or traditional discussion boards in the Internet. Instead, arguments are
raised about a particular topic, and citizens can then agree or disagree with
those. Participants can also raise new arguments for others to vote. Thanks to
real-time analytics, participating citizens can also observe the formation of
agreeing or disagreeing opinion groups. In contrast to the polarizing
discussions and algorithms on social media, these features should help at
reaching a consensus about a policy topic. Although the functionality may be
slightly different in other implementations, such as
Decidim~\cite{Barandiaran24}\footnote{~\url{https://decidim.org/}} and CONSUL
DEMOCRACY\footnote{~\url{https://consuldemocracy.org/}}, the basic ideas are
still typically shared across different citizen platform implementations. Also
tailored citizen platforms have been proposed~\cite{Vicens18}. Against this
backdrop, it might be possible to develop a custom platform specifically for
requirements engineering.

\section{Citizen Platforms and Requirements Engineering}\label{sec: citizen platforms re}

The functionalities of the Polis platform help at presenting a few high-level
design ideas for such a platform. The iterative requirements engineering
practices typically used in agile software development~\cite{Ramesh10} can be
used to frame the ideas. Thus, a requirements engineering process on a citizen
platform might involve five phases: (1)~a preparation phase, (2) an initial
requirements engineering phase, (3)~a prioritization phase, (4) a development
phase, and (5) an acceptance phase.

Of these, the first phase would presumably be critical for successfully
completing the later phases. Among other things, requirements engineers,
possibly together with a product owner, would likely need to carefully write
instructions and guidelines about a given large socio-technical system
envisioned. These should probably also include high-level materials on matters
such as user stories. Alternatively, it might be possible to arrange a
preliminary offline or online lecture about these topics Some citizen platform
implementations, such as the noted Decidim~\cite{Barandiaran24}, have also been
designed with education and pedagogy in mind. In addition to a potential risk of
failing at delivering adequate instructions, the phase involves also a difficult
question about how to recruit and invite citizens and stakeholders to a
platform. It remains generally unclear how many citizens and stakeholders should
be invited, and how representative they would be in terms of a whole
society~\cite{Secinaro21, Simonofski21}. While techniques from survey research
might be used to gain representativeness for the recruitment invitations, it
might well be that the resulting participants would be biased toward
stakeholders on one hand and those citizens who are already engaged on the
other. Such a democratic (digital) divide is well-known in the
literature~\cite{Nam12}. The potential bias would be particularly severe in
case citizens for whom a system is geared are missing. For instance, it would
seem oblivious to develop a public sector system for helping people to find
employment in case unemployed people are missing.

In the second phase the citizens and stakeholders invited would engage on a
platform by discussing a socio-technical system planned, raising high-level
requirements as arguments, contemplating these with the help of analytics and a
facilitating requirements engineer, and finally, if necessary, voting on the
most important or preferable requirements. Also this phase as well as later
gatherings on a platform contain their own risks.

As face-to-face requirements elicitation is preferred in agile
development~\cite{Ramesh10}, digitalized platform-based elicitation would likely
require new skills and techniques from software engineers. A similar point
applies to participating citizens and stakeholders. Analogously to the so-called
co-design practices, it might be that participants with technological or
domain-specific knowledge would dominate those without such knowledge; so-called
end-users would be overruled by so-called power users who are closer to being
developers themselves~\cite{Fischer20, Secinaro21}. A similar point applies with
respect to powerful stakeholders and interest groups.

It might also be that ordinary citizens would have difficulties to articulate
requirements. This point underlines the importance of the prior instructions and
guidelines. Analogously to focus groups~\cite{Kontio08}, the role of a
facilitating and moderating requirements engineer is likely highly important
too~\cite{Barandiaran24, Dalpiaz16}. If there are difficulties to generate
requirements, or if a discussion gets stuck, he or she should have the skills to
put the process again on the right track. Given the public sector context, it
might also be that a discussion gets too heated; hence, he or she should also
know how to moderate a discussion, possibly by using a platform's moderation
tools. Because citizen platforms should engage also citizens who possibly speak
different languages~\cite{Simonofski21}, language barriers might pose an
additional risk, including with respect to a facilitator.

After the initial gathering on a platform, requirements engineers and other
developers would engage during the prioritization phase by evaluating
particularly the feasibility of the requirements put forward, ranking these, and
sketching design ideas for implementing these. Given the context of large public
sector socio-technical systems, the evaluation process should involve also
public sector representatives with expertise in financial, legal, and related
matters. Then, the prioritized requirements would be either delivered to the
citizens and stakeholders directly or a second gathering would be arranged on a
platform for the citizens and stakeholders to evaluate the prioritized
requirements, possibly again voting on these and their ranking. There are risks
affecting also this phase, at least in the form sketched. Because civil servants
would be likely needed in the feasibility evaluations, the process sketched
would imply a double burden for requirements engineering. In addition, it might
be that not the citizens and stakeholders on a platform but the civil servants
would fail to reach a consensus. Particularly financial matters might be
difficult to agree upon. A further point is that legal requirements are not easy
to specify~\cite{Hjerppe19RE}. Given the public sector context, financial
uncertainties and legal obstacles might even lead to a failure of the
prioritization phase altogether. To counter these risks, it might be possible to
use a different platform instance for this auxiliary group of participants;
after all, they are---or should be---experts in deliberation already. Some
citizen platform implementations \cite{Barandiaran24} offer also tools for
budgeting and project financing.

The development and acceptance phases would then proceed as is common in agile
software development. After an iteration of development, a gathering might again
be arranged on a platform for the citizens and stakeholders to evaluate and
discuss a given system being developed. Either voting or informal
consensus-seeking might be again used to evaluate satisfaction with the system
together with possibly new requirements that may emerge in the
discussions. Given that the amount of participating citizens and stakeholders
would presumably be much larger than in conventional software projects, it might
be reasonable that a single gathering on a platform would cover multiple
development iterations. Another, alternative option might be to impose fixed
time limits for the deliberations. Regardless, during the final acceptance phase
a finished system would be evaluated particularly against the initial and
subsequent requirements. While voting or informal deliberation might be again
used, it might be also possible to develop a citizen platform in a way that
supports the ideas of acceptance testing.

Furthermore, it may be that plain voting, while necessary for a democracy, is
not particularly well-suited for requirements engineering, despite it sometimes
appearing in the literature~\cite{Groen17, Ramesh92}. This point applies to
co-design and co-creation as well. A~memorable illustration of the point
appeared in the Simpsons cartoon series; in an episode Homer had a chance to
design a new car, and, expectedly, it failed miserably because a lot of unneeded
and undesirable features were
implemented.\footnote{~\url{https://simpsons.fandom.com/wiki/The_Homer}} In
other words, so-called feature creep too is a risk for large-scale participatory
requirements engineering on a citizen platform. Given that deliberation and
consensus-seeking are continuous endeavors, the risk further entails the
commonplace problem of changing requirements, which have often been seen as a
major cause for feature creep~\cite{Jones96}. Due to these and related reasons,
it may also be that citizen platforms are a poor solution to develop innovative
systems. Though, unlike in product development, it must be added that
innovativeness is not necessarily a primary or even a top-ranked criterion for
the development of large socio-technical systems in the public sector context.

Against these backdrops, nevertheless, a citizen platform for requirements
engineering may need more complex voting techniques that take different
feasibility and related criteria into account, including financial, legal, and
other constraints that are important for requirements engineering and software
design~\cite{Budgen03, Hjerppe19RE}. It might also be reasonable to prototype
and evaluate so-called downvoting functionality. Even with such techniques and
functionalities, a key point should be kept in mind: the ``better the epistemic
processes of deliberation \textit{before} voting---including, where needed, the
transmission of relevant expert knowledge to the broader public---the more
likely it is that votes will be epistemically efficient in the sense that
individuals are well informed about the subject matters at
stake''~\cite[p.~71]{Herzog24}. A further important point about voting is
related to security and privacy. In case sensitive matters are voted upon,
confidentiality should be guaranteed analogously to electoral secrecy in
conventional democratic~elections.

As has been also pointed out in a related context~\cite{Groen17}, a citizen
platform for requirements engineering may also benefit from customized
algorithms, including natural language processing methods tailored for
requirements specifically. These methods are also a thriving research domain in
requirements engineering~\cite{ZhaoAlhoshan21}. Additionally, techniques such as
gamification might be used~\cite{Dalpiaz16}, although the so-called nudging
involved in these techniques arguably aligns poorly with the fundamental ideals
of deliberative democracy. In other words, even in the presence of tools and
algorithms, the central design idea should be that citizens and stakeholders can
freely, without manipulation or other push factors, reach a consensus about
requirements in their own terms through informed discussions. For such
deliberation to work properly, openness and transparency are particularly
important factors in the public sector context~\cite{Nam12}. A lack of these has
often been seen as a major problem for the today's large commercial social media
platforms~\cite{Ruohonen24FM}. To this end, it suffices to end the discussion by
emphasizing a risk of replicating existing problems insofar as deliberative
democracy is considered.

\section{Conclusion}\label{sec: conclusion}

This short vision paper presented preliminary but novel ideas about a potential
use of citizen platforms for requirements engineering of large public sector
socio-technical systems. In contrast to the crowd-sourcing ideas already
presented in the requirements engineering literature, these platforms are
essentially about consensus-seeking through deliberation, including on
potentially controversial or otherwise hot political topics.

By and large, these platforms have been previously used for politics and
policies, not implementations based on the policies enacted. In terms of
governance of large public sector socio-technical systems, both citizens and
stakeholders exert pressure upon political decision-makers who enact new
technology regulations~\cite{Musil20}. By implication, in order to reach the
full potential of civic engagement, citizen platforms could (or should) be used
for both the formulation of technology policies and the implementation of
conforming software solutions. There is a lot to improve also on the traditional
policy side. For instance, in the European Union (EU), from which most
technology policies nowadays come from in Europe, the associated policy
consultations can gather thousands of responses~\cite{Ruohonen24FM}, and it
remains unclear how well the voices of EU citizens are heard in the face of
heavy lobbying by powerful industry groups. In addition, policy consultations
have obvious limitations in that these are largely non-deliberative and
particularly non-iterative. Thus, maybe there might be room to pilot citizen
platforms also in the EU's future technology policy-making. Having said, it
remains generally unclear how the potential risk of hijacking or otherwise
dominating a deliberation process could be addressed. This risk reinforces the
earlier points raised about representativeness, facilitation, knowledge-delivery
and epistemology in general, and even content moderation.

Piloting a citizen platform for requirements engineering would be a good way to
continue in research. A pilot study should shed more light on the potential
advantages and disadvantages, including the risks discussed. In particular,
better understanding is required on whether and how well the principles of
deliberative democracy apply to requirements engineering. In this regard, there
are two fundamental evaluation dimensions to consider. The first is the
outcome-side~\cite{Nam12}, that is, in the present context the quality and other
related properties of a socio-technical system designed and implemented. Such an
evaluation requires that a pilot should involve a design of a real-world
system. The second is the platform-side, including with respect to the risk
discussed. On this side, an evaluation should try to also reflect upon a risk of
potential \textit{participatory washing}, meaning the hiding of existing
problems in democracies underneath platforms and technology in
general~\cite{Barandiaran24}. Then, if a pilot study indicates that citizen
platforms are poorly suited for requirements engineering, it might be reasonable
to contemplate alternatives. An example would be public procurement of large
socio-technical systems; here, a citizen platform might be used to evaluate
tendering in terms of acceptability and desirability among citizens. As was
discussed, a further research avenue opens on the implementation side; it might
be desirable to implement a new citizen platform specifically for requirements
engineering. Though, in order to avoid wasting resources, the applicability of
the existing citizen platform implementations should be evaluated first.

\section*{Disclosure of Interests}

There are neither competing nor conflicting interests.

\bibliographystyle{splncs03}
%\bibliography{se}

\end{document}